\documentclass[12pt]{article}
\usepackage{amssymb}
\usepackage{color,graphicx}
\usepackage{amsmath}

\usepackage{hyperref}

\begin{document}

\renewcommand{\abstractname}{\hfill}

\newpage
\pagenumbering{arabic}
\begin{center}
\bf
Yu. V. Pavlov\,\footnote{\,Institute for Problems in Mechanical Engineering
of Russian Academy of Sciences,
Saint-Petersburg, Russia.\, e-mail:\, yuri.pavlov@mail.ru \ (corresponding author).},
V. P. Vandeev\,\footnote{\,Petersburg Nuclear Physics Institute
named by B. P. Konstantinov of National Research Centre ``Kurchatov
Institute'', Gatchina, Leningrad Region, Russia,\,
e-mail:\, vandeev\_vp@pnpi.nrcki.ru}
\end{center}

\begin{center}
\large \bf \uppercase{Deviation of Trajectories in Nonrelativistic\\[4pt]
and Relativistic Cases and Shirokov Effect}
\end{center}

\begin{abstract}
    We compare the deviations of circular orbits in the nonrelativistic and relativistic cases.
    General solutions of the deviation equations are obtained.
    We find the potentials for the nonrelativistic case and the metric components
in general relativity for which trajectories close to circular ones are closed curves.
    Explicit expressions for the pericenter shift of an orbit close to
a circular one in a static spherically symmetric spacetime are obtained.
    Estimates of the cosmological constant effect on the orbit pericenter shift are given.
    Finally, we find the spherically symmetric metric in which the Shirokov effect is absent.
\end{abstract}


\section{\normalsize Introduction}

\hspace*{\parindent}
    In studying perturbations of the motion trajectories for bodies in
general relativity, a frequency difference was found between the orthogonal
oscillations of test bodies in a circular orbit in the Schwarzschild
field~\cite{Shirokov73}.
    In classical mechanics, test bodies rotating in a circular orbit around
the center of gravity oscillate in all directions with the same frequency,
which is equal to the orbital rotation frequency.
    Such a difference in oscillation frequencies in the theory of relativity
is called the Shirokov effect.
    This effect has been studied in a number of papers
(see, e.g.,~\cite{Greenberg74}---\cite{Laemmerzahl16}).

    The purpose of our paper is to compare the description of circular orbit
deviations for arbitrary centrally symmetric potentials in classical mechanics
with that for static spherically symmetric spaces in general relativity.
    To this end, we obtain and study the general solutions of the deviation
equations in both cases.
    We find the spacetime metrics in which the geodesic deviations are similar
to the deviations from circular trajectories in Newtonian theory
for potentials of a special form.
    We consider the cases of closed trajectories, as well as the limiting
stable and unstable orbits.
    In our paper, we also answer the question of the existence of a spacetime
for which there is no Shirokov effect, i.e., where the deviations oscillate
with the same frequency in all directions.

    The phenomenon of a mismatch in oscillation frequencies is related to
the pericenter shift of the perturbed orbit.
    In this work, we give approximate estimates of the influence of
the cosmological constant on the pericenter shift of an orbit close to
a circular one in the Schwarzschild field, and analyze the possibilities
of observing this effect.

    The paper is organized as follows.
    In Sec.\,\ref{Sec2}, the deviations of circular orbits in classical mechanics
for centrally symmetric potentials are considered, general solutions of
the deviation equations are obtained, and closure conditions for trajectories
are established.
    In Sec.\,\ref{Sec3}, the deviation of a circular trajectory
in general relativity in a static spherically symmetric spacetime is considered,
and the explicit form of the general solutions of the deviation equations
is obtained.
    In Sec.\,\ref{Sec4}, the relations between the orbit pericenter shift and
the difference in the oscillation frequencies of perturbed trajectories
are given, and estimates of the influence of the cosmological constant on
the pericenter shift are obtained.
    In Sec.~\ref{Sec5}, we find examples of spaces where the deviations of
circular geodesics are closed curves.
    In Sec.~\ref{Sec6}, a spherically symmetric spacetime is found in which
the deviations of circular geodesics behave similarly to the case of
a potential inversely proportional to the distance in classical mechanics,
i.e., to the limiting case of stable orbits.
    In Sec.\,\ref{Sec7}, an example of a spacetime without the Shirokov effect
and the energy-momentum tensor for the matter generating it are found.
    The obtained results are summarized in Sec.~\ref{secZakl}.

\vspace{3mm}
{\centering \section{\normalsize Non-relativistic case
\label{Sec2}}}

\hspace{\parindent}
   If $ U(\mathbf{r}) $ is the potential energy of a particle of mass~$m$
in a given conservative force field, the equation of motion in
Cartesian coordinates is given by
    \begin{equation}    \label{u2}
\ddot{x}^k = - \partial^k \varphi(\mathbf{r}).
\end{equation}
    where the dot denotes the time derivative, and
$\varphi(\mathbf{r}) = U(\mathbf{r}) /m$.

    Let~$x^k(t)$ be the trajectory of some particle, and let $X^k(t)$
be another trajectory of the particle close to the given one within
the same potential
$ \varphi(\mathbf{r}) $:
    \begin{equation}    \label{u3}
X^k(t) = x^k(t)+ \eta^k(t),
\end{equation}
    where $\eta^k$ are the coordinates of the deviation vector.
    From the equations of motion in Cartesian coordinates, we obtain
    \begin{equation}    \label{u4}
\ddot{X}^k = \ddot{x}^k + \ddot{\eta}^k =
- \partial^k \varphi(\mathbf{x}+ {\bold{ \eta } }) .
\end{equation}
    Expanding the potential to order $ O(\eta^2)$
    \begin{equation}    \label{u5}
\varphi(\mathbf{x}+ {\bold{ \eta } }) = \varphi(\mathbf{x}) +
\eta^n \partial_n \varphi (\mathbf{x}) + O(\eta^2)
\end{equation}
    (where summation over repeated indices is implied) and taking into account
Eq.~(\ref{u2}), we derive the deviation equation for the trajectory in
Cartesian coordinates from Eq.~(\ref{u4})
    \begin{equation}    \label{u6}
\ddot{\eta}^k = - \eta^n \partial^k \partial_n \varphi(\mathbf{x}) .
\end{equation}
    For a central field $\varphi(\mathbf{r})=\varphi(r)$,
deviation equation~(\ref{u6}) takes the form
    \begin{equation}    \label{ndev}
\frac{d^2 \eta^k}{d t^2} = - \eta^k \frac{\varphi'}{r} - x^k \eta^i x^i
\left( \frac{\varphi''}{r^2} - \frac{\varphi'}{r^3} \right),
\end{equation}
    where primes denote derivatives with respect to~$r$.

    We consider the case of circular orbits.
    The particle energy in a central field equals~\cite{LL_I}
    \begin{equation}    \label{effE}
E = \frac{m \dot{r}^2}{2} + m \varphi_{\rm eff}(r),
\end{equation}
    where
    \begin{equation}    \label{effp}
\varphi_{\rm eff}(r) = \varphi(r) + \frac{L^2}{2 m^2 r^2}
\end{equation}
    is the effective potential, and $L$ is the angular momentum of the particle
relative to the center of gravity $r=0$.
    The condition for the existence of circular orbits follows from~(\ref{effp}):
    \begin{equation}    \label{effko}
\varphi_{\rm eff} (r) = \frac{E}{m}, \ \ \ \
\varphi_{\rm eff}'(r)= \varphi'(r) - \frac{L^2}{m^2 r^3} = 0.
\end{equation}
    We note that, according to~(\ref{effko}), one has for a circular orbit
    \begin{equation}    \label{effkop}
\varphi'(r) = \frac{L^2}{m^2 r^3} >0.
\end{equation}
    The angular frequency $\omega = L/(m r^2)$ of rotation in
the circular orbit equals
    \begin{equation}    \label{u7}
\omega = \sqrt{ \frac{\varphi'(r)}{r}}.
\end{equation}
    The circular orbit with a given radius $r$ is stable if
    \begin{equation}    \label{effu}
\varphi_{\rm eff}''(r) > 0.
\end{equation}
    From~(\ref{effko}), we obtain the second derivative of the effective
potential in the circular orbit with the given $r$:
    \begin{equation}    \label{effdu}
\varphi_{\rm eff}''(r) = \varphi'' + 3 \frac{\varphi'}{r}.
\end{equation}
    Therefore, in a stable circular orbit,
    \begin{equation}    \label{effduk}
\varphi'' + 3 \frac{\varphi'}{r} > 0.
\end{equation}

    Next, we write the deviation equation of the trajectory with respect to
the circular orbit.
    We use spherical coordinates:
    \begin{equation}    \label{u8}
x^1 = r \sin \theta \cos \phi, \ \ \
x^2 = r \sin \theta \sin \phi, \ \ \
x^3 = r \cos \theta .
\end{equation}
    The Cartesian basis vectors $  \mathbf{i},  \mathbf{j},  \mathbf{k} $
and the unit vectors $\mathbf{e}_r, \mathbf{e}_\theta, \mathbf{e}_\phi $
in the directions $ r, \theta, \phi $ are related as follows:
    \begin{equation}    \label{u9}
\left\{
\begin{array}{l}
\mathbf{e}_r = \mathbf{i} \cos \phi \sin \theta +
\mathbf{j} \sin \phi \sin \theta + \mathbf{k} \cos \theta, \\[5pt]
\mathbf{e}_\theta = \mathbf{i} \cos \phi \cos \theta +
\mathbf{j} \sin \phi \cos \theta - \mathbf{k} \sin \theta, \\[3pt]
\mathbf{e}_\phi = - \mathbf{i} \sin \phi + \mathbf{j} \cos \phi.
\end{array}
\right.
\end{equation}
    Therefore, for the coordinates of an arbitrary vector $\mathbf{V}$
in the Cartesian and spherical bases, we obtain
    \begin{equation}    \label{u11}
\left\{
\begin{array}{l}
V^1 = V^r \cos \phi \sin \theta + V^\theta \cos \phi \cos \theta
- V^\phi \sin \phi, \\[5pt]
V^2 = V^r \sin \phi \sin \theta + V^\theta \sin \phi \cos \theta
+ V^\phi \cos \phi, \\[3pt]
V^3 = V^r \cos \theta - V^\theta \sin \theta .
\end{array}
\right.
\end{equation}

    For the case of deviation with respect to a circular orbit of radius~$r$
lying in the plane $\theta = \pi/2$, we have
    \begin{equation}    \label{u12}
x^1 = r \cos \phi, \ \ \ x^2 = r \sin \phi, \ \ \ x^3 = 0,
\end{equation}
    \begin{equation}    \label{u13}
\eta^1 = \eta^r \cos \phi - \eta^\phi \sin \phi, \ \ \
\eta^2 = \eta^r \sin \phi + \eta^\phi \cos \phi, \ \ \
\eta^3 = - \eta^\theta.
\end{equation}
    Substituting Eqs.~(\ref{u12}) and (\ref{u13}) into deviation
equation~(\ref{ndev}), we obtain, respectively:
    \begin{equation}  \label{ndev1}
\frac{d^2 \eta^r}{d t^2} \cos \phi - 2 \frac{d \eta^r}{d t} \omega \sin \phi -
\frac{d^2 \eta^\phi}{d t^2} \sin \phi - 2 \frac{d \eta^\phi}{d t} \omega
\cos \phi =  - \eta^r \left( \varphi'' - \frac{\varphi'}{r} \right) \cos \phi,
\end{equation}
    \begin{equation}  \label{ndev2}
\frac{d^2 \eta^r}{d t^2} \sin \phi + 2 \frac{d \eta^r}{d t} \omega \cos \phi
+ \frac{d^2 \eta^\phi}{d t^2} \cos \phi - 2 \frac{d \eta^\phi}{d t} \omega
\sin \phi = - \eta^r \left( \varphi'' - \frac{\varphi'}{r} \right) \sin \phi .
\end{equation}
    \begin{equation}    \label{ndev3}
\frac{d^2 \eta^\theta}{d t^2} = - \eta^\theta \frac{\varphi'}{r} .
\end{equation}

    By virtue of~(\ref{u7}), Eq.~(\ref{ndev3}) takes the form
    \begin{equation}    \label{ndev3du}
\frac{d^2 \eta^\theta}{d t^2} + \omega^2 \eta^\theta =0
\end{equation}
    and has the general solution
    \begin{equation}    \label{ndev3d}
\eta^\theta = C_{1 \theta} \sin \omega t + C_{2 \theta} \cos \omega t,
\end{equation}
    where $C_{1 \theta}$ and $ C_{2 \theta}$ are arbitrary constants.

    Therefore, the oscillations of the deviation vector in the direction of
the azimuthal angle~$\theta$ have the same frequency as
the orbital rotation of the body.
    This result has the following explanation.
    The orbital plane remains fixed in space.
    Therefore, the oscillating body intersects the orbital plane of another
close body exactly twice during one period, and the relative shift in the
direction perpendicular to this plane oscillates at the body's orbital frequency.

    Multiplying Eq.~(\ref{ndev1}) by $\cos \phi $ and Eq.~(\ref{ndev2})
by $\sin \phi $, and adding them, we obtain
    \begin{equation}    \label{ndev1d}
\frac{d^2 \eta^r}{d t^2} + \eta^r \left( \varphi'' - \frac{\varphi'}{r} \right)
- 2 \omega \frac{d \eta^\phi }{d t} = 0.
\end{equation}
    Multiplying Eq.~(\ref{ndev1}) by ($ - \sin \phi $) and Eq.~(\ref{ndev2})
by $\cos \phi $, and adding them, we obtain
    \begin{equation}    \label{ndev2d}
\frac{d^2 \eta^\phi}{d t^2} + 2 \omega \frac{d \eta^r }{d t} = 0.
\end{equation}

    To find the general solution of the system of equations~(\ref{ndev1d})
and~(\ref{ndev2d}), we differentiate Eq.~(\ref{ndev1d}) with
respect to time~$t$ and substitute $ d^2 \eta^\phi / d t^2 $
from Eq.~(\ref{ndev2d}), obtaining
    \begin{equation}    \label{ndev1dd}
\frac{d^3 \eta^r}{d t^3} + \frac{d \eta^r }{d t}
\left( \varphi'' + 3 \frac{\varphi'}{r} \right) = 0.
\end{equation}
    Let us introduce the notation
    \begin{equation}    \label{Omega}
\Omega^2 = \varphi'' + 3 \frac{\varphi'}{r} .
\end{equation}
    According to Eq.~(\ref{effdu}), the quantity $ \Omega^2$ in a circular
orbit coincides with the second derivative of the effective potential and,
by Eq.~(\ref{effduk}), it is positive for stable circular orbits.
    The value $ \Omega^2=0$ holds for the potentials $\varphi = \alpha /r^2$,
in which circular orbits exist if $\alpha <0$ for the energy and
angular momentum values
    \begin{equation}    \label{n32k}
E=0, \ \ \  L^2 = - 2 \alpha m^2,
\end{equation}
    at any radial distance from the center of gravity.

    The general solution of Eq.~(\ref{ndev1dd}) is given by
    \begin{equation}    \label{OR1}
\eta^r =
\left\{
\begin{array}{ll}
C_1 \sin \Omega t + C_2 \cos \Omega t + C_3,  & \Omega^2 > 0, \\[4pt]
C_1 t^2 + C_2 t + C_3,  & \Omega^2 = 0, \\[4pt]
C_1 \sinh |\Omega| t + C_2 \cosh |\Omega| t + C_3,  & \Omega^2 <0 ,
\end{array}
\right.
\end{equation}
    where $C_1$, $C_2$, and $C_3$ are arbitrary constants.

    Let us obtain the general solution for $\eta^\phi$.
    It follows from Eq.~(\ref{ndev2d}) that
    \begin{equation}    \label{ndev26s}
\frac{d \eta^\phi}{d t} = - 2 \omega ( \eta^r - C_3) + C_5,
\end{equation}
    where $C_5$ is a constant.
    Substitution of this expression into Eq.~(\ref{ndev1d}) yields
    \begin{equation}    \label{svyaz}
C_5 = \frac{\Omega^2 - 4 \omega^2}{2 \omega} C_3.
\end{equation}
    Integrating Eq.~(\ref{ndev26s}), we obtain
    \begin{equation}    \label{OR2gen}
\eta^\phi =
\left\{
\begin{array}{ll}
2 \frac{\omega}{\Omega} C_1 \cos \Omega t - 2 \frac{\omega}{\Omega} C_2
\sin \Omega t + \frac{\Omega^2 - 4 \omega^2}{2 \omega} C_3 t + C_4, & \Omega^2 > 0, \\[5pt]
- \frac{2}{3} \omega C_1 t^3 - \omega C_2 t^2  - 2 \omega C_3 t + C_4, & \Omega^2 = 0, \\[5pt]
-2 \frac{\omega}{|\Omega|} C_1 \cosh |\Omega| t -
2 \frac{\omega}{|\Omega|} C_2 \sinh |\Omega| t + \frac{\Omega^2 -
4 \omega^2}{2 \omega} C_3 t + C_4, & \Omega^2 <0 ,
\end{array}
\right.
\end{equation}
    where $C_4$ is an arbitrary constant.

    In the case of the gravitational field of a point mass~$M$,
the frequencies $\omega$ and $\Omega$ coincide:
    \begin{equation}    \label{tm}
\varphi = - \frac{GM}{r} \ \ \Rightarrow \ \
\Omega = \omega = \sqrt{\frac{GM}{r^3} },
\end{equation}
    where $G$ is the gravitational constant.
    The circular orbits in such a potential exist at any value of~$r$,
and all of them are stable.

    For all other potentials
(potentials which differ by an additive constant are considered identical),
the oscillation frequencies in the $\theta$-direction (frequency $\omega$)
and in the $ r, \phi $-directions (frequency $\Omega$) are different:
    \begin{equation}    \label{omom}
\Omega = \omega \ \ \Leftrightarrow \ \ \varphi'' + 2 \frac{\varphi'}{r} =0,
\ \ \varphi' >0 \ \ \Leftrightarrow \ \ \varphi = - \frac{C'_1}{r} + C'_2 ,
\end{equation}
    where $C'_1>0$ and $C'_2$ are arbitrary constants.

    Thus, for central potentials, the deviation oscillations for a circular
trajectory in the nonrelativistic case in the $\theta$-direction always have
the same frequency as the circular rotation frequency~$ \omega$.
    The oscillations in the $r$, $\phi$-directions have the frequency~$\Omega$
coinciding with~$ \omega$ only in the case of attraction with a potential $\sim 1/r$.

     We consider the question of the closedness of trajectories
that are close to circular ones.
    According to Bertrand's theorem~\cite{Bertrand1873},
for the motion of a particle in a central potential field,
all finite trajectories are closed only for the potentials
$U(r) = -\alpha/ r$ $ (\alpha > 0)$ and $U(r) = k r^2 $ $(k > 0)$.

     In the case of trajectories deviated from circular ones,
the closedness condition is the commensurability of the oscillation frequencies
$\omega$ and $\Omega$ in different directions, i.e., the existence of natural
numbers $m$ and $n$ such that
    \begin{equation}    \label{sch1}
m \omega = n \Omega .
\end{equation}
    This condition reduces to solving the equation for the potential
    \begin{equation}    \label{sch2}
\phi'' + \frac{3 \phi'}{r} = \frac{m^2}{n^2} \frac{\phi'}{r}
\end{equation}
    under the condition $ \phi' > 0$
(the existence condition for circular orbits).
    The solutions of~(\ref{sch2}) are potentials of the form
    \begin{equation}    \label{sch4}
\phi= \alpha\, r^{\left( \frac{m^2}{n^2}-2 \right)} + \tilde{C},
\ \ \ \alpha \left( \frac{m^2}{n^2}-2 \right) >0,
\end{equation}
    where $\alpha$ and $\tilde{C}$ are constants.
    For $m/n=1$ and $m/n=2$ formula~(\ref{sch4}) reproduces the potentials in
Bertrand's theorem.
    However, for small deviations from circular motion, trajectories become
closed for an infinite number of other potentials~(\ref{sch4})
as well, corresponding to different fractional numbers $m/n$.

\vspace{3mm}
{\centering \section{\normalsize Deviation of circular orbits in
general relativity \label{Sec3}}}

\hspace{\parindent}
    According to general relativity, the free motion of a particle
in a gravitational field follows the geodesic line
         \begin{equation}
\frac{d u^i}{d s} + \Gamma^{\, i}_{\, kl} u^k u^l =0,
\label{geu}
\end{equation}
    where $\Gamma^{\, i}_{\, kl} $ are the Christoffel symbols of the
corresponding curved spacetime, and $u^i$ is the 4-velocity of the particle.
    Hereafter, the Latin indices take values $0,1,2,3$,
and summation over repeated indices is implied.

    The deviation of two particles moving along close geodesic lines is
described by the equation of geodesic deviation~\cite{LL_II}
    \begin{equation}    \label{DGU}
\frac{D^2 \eta^i}{d s^2}= R^i_{klm} u^k u^m \eta^l,
\end{equation}
    where
    \begin{equation}
R^{\, i}_{\ klm} = \partial_{\,l}  \Gamma^{\, i}_{\, km} -
\partial_m  \Gamma^{\, i}_{\, kl} +
\Gamma^{\, i}_{\, nl} \Gamma^{\, n}_{\, km} -
\Gamma^{\, i}_{\, nm} \Gamma^{\, n}_{\, kl}
\label{Rijk}
\end{equation}
    is the curvature tensor,
    \begin{equation}    \label{DGU01}
\frac{D \eta^i}{d s}= (\nabla_k \eta^i) u^k,
\end{equation}
    and $\nabla_k $ is the covariant derivative.
    Calculating the covariant derivatives and taking into account Eq.~(\ref{geu}),
we obtain the equation of geodesic deviation in the form
    \begin{equation}
\frac{d^2 \eta^i}{d s^2} + 2 \Gamma^{\, i}_{\, kl} u^k \frac{d \eta^l}{d s} +
\frac{\partial \Gamma^{\, i}_{\, kl}}{\partial x^m} u^k u^l \eta^m = 0.
\label{D2nsNF}
\end{equation}

    Next, we consider the case of a static spherically symmetric spacetime
with the interval
    \begin{equation}    \label{msf}
d s^2 = f(r)\, c^2 d t^2 - \frac{d r^2}{B(r)} - r^2 \left( d \theta^2 +
\sin^2 \theta \, d \phi^2 \right).
\end{equation}
    We denote the coordinates as $x^0 = c t$, $x^1 = r$, $x^2 = \theta$, and
$x^3= \phi$.
    The nonzero Christoffel symbols for this case are as follows~\cite{LL_II}:
    \begin{eqnarray} \nonumber
\Gamma_{10}^{\,0} = \frac{f'}{2 f}, \ \ \
\Gamma_{00}^{\,1} = \frac{f'}{2} B, \ \ \
\Gamma_{11}^{\,1} = - \frac{B'}{2 B}, \ \ \
\Gamma_{22}^{\,1} = - r B, \ \ \
\Gamma_{33}^{\,1} = - r B \sin^2 \theta , \\
\Gamma_{33}^{2} = - \frac{ \sin (2 \theta) }{2}, \ \ \
\Gamma_{12}^{\,2} = \Gamma_{13}^{\,3} = \frac{1}{r}, \ \ \
\Gamma_{23}^{\,3} = \frac{\cos \theta}{\sin \theta }.
\label{Gamma}
\end{eqnarray}

    Let us consider the free motion of a massive particle in a circular orbit
in the gravitational field~(\ref{msf}).
    In circular orbits, $u^1=0$.
    We select the frame of reference such that the plane of the particle orbit
corresponds to the value~$\theta = \pi/2$.
    Then the corresponding component of the 4-velocity is $u^2=0$.
    To find the components $u^0$ and $u^3$, we divide both sides of
Eq.~(\ref{msf}) by $ds^2$:
    \begin{equation}    \label{msf1}
f (u^0)^2 - r^2 (u^3)^2 = 1 .
\end{equation}
    From geodesic equation~(\ref{geu}) for $i=1$, we obtain
    \begin{equation}    \label{msf2}
\frac{f' B}{2} (u^0)^2 - r B (u^3)^2 = 0 .
\end{equation}
    Solving system~(\ref{msf1}) and (\ref{msf2}), we obtain for circular orbits
    \begin{equation}    \label{kr0}
(u^0)^2 = \frac{2}{2 f - r f'},
\end{equation}
    \begin{equation}    \label{kr1}
(u^3)^2 = \frac{f'}{2 r f - r^2 f'}.
\end{equation}

    Substituting Eqs.~(\ref{Gamma}) with $ \theta= \pi/2 $ and the found values
of the 4-velocity into Eq.~(\ref{D2nsNF}),
we obtain the following system of equations for the deviation of
circular orbits in a static spherically symmetric spacetime:
    \begin{equation}    \label{shx0}
\frac{d^2 \eta^0}{d s^2} + A \frac{d \eta^1}{d s} = 0,
\end{equation}
    \begin{equation}    \label{shx1}
\frac{d^2 \eta^1}{d s^2} + B_1 \frac{d \eta^0}{d s} +
B_2 \frac{d \eta^3}{d s} + B_3 \eta^1 = 0,
\end{equation}
    \begin{equation}    \label{shx2}
\frac{d^2 \eta^2}{d s^2} + C \eta^2 = 0,
\ \ \ \ C = (u^3)^2,
\end{equation}
    \begin{equation}    \label{shx3}
\frac{d^2 \eta^3}{d s^2} + D \frac{d \eta^1}{d s} = 0,
\end{equation}
    where
    \begin{equation}    \label{shx1d}
A = \frac{f'}{f} u^0, \ \ \
B_1 = f' B u^0, \ \ \ B_2 = - 2 r B u^3 , \ \ \
B_3 = \frac{(u^0)^2}{2} (f' B)' - (u^3 )^2 (r B)', \ \ \ D = \frac{2}{r} (u^3).
\end{equation}

    The equation for the $\theta$-component does not depend on the other
equations of the system and has the general solution
    \begin{equation}    \label{shx2Res}
\eta^2 = \theta =  C_1 \cos \omega \tau + C_2 \sin \omega \tau,
\end{equation}
    where $\tau = s/c$ is the proper time of the rotating particle, and
$C_1$ and $C_2$ are arbitrary constants.
    Thus, as in the nonrelativistic case (see Eq.~(\ref{ndev3d})),
the motion of the $\theta$-component of deviation represents a harmonic
oscillation with the angular frequency $\omega $ equal to
the rotation frequency in the circular orbit:
    \begin{equation}    \label{sh1}
\omega^2 = (c u^3)^2 = \frac{c^2 f'}{2 r f - r^2 f'}.
\end{equation}

     Equations~(\ref{shx0}), (\ref{shx1}), and (\ref{shx3}) for the other
components represent a system of linear homogeneous differential equations.
    The general solution of this linear system must contain six arbitrary constants.
We look for a solution of this system in the form
    \begin{equation}    \label{shxRS}
\eta^0 = \eta^0_{(0)} \exp (\pm i \Omega s/c), \ \ \ \
\eta^1 = \eta^1_{(0)} \exp (\pm i \Omega s/c), \ \ \ \
\eta^3 = \eta^3_{(0)} \exp (\pm i \Omega s/c).
\end{equation}
    Then Eqs.~(\ref{shx0}), (\ref{shx1}), and (\ref{shx3}) reduce to a system
of linear algebraic equations:
    \begin{equation}    \label{tmSO}
\left\{
\begin{array}{l}
- \Omega^2 \eta^0_{(0)} + i \Omega c A \eta^1_{(0)} = 0, \\[5pt]
i \Omega c B_1 \eta^0_{(0)} +(c^2 B_3 - \Omega^2) \eta^1_{(0)} +
i \Omega c B_2 \eta^3_{(0)} =0, \\[5pt]
i \Omega c D \eta^1_{(0)} - \Omega^2 \eta^3_{(0)} = 0 .
\end{array}
\right.
\end{equation}
    For the existence of nonzero solutions, it is necessary and sufficient
that the determinant of this system equal zero:
    \begin{equation}    \label{tmS1}
\Omega^4 (B_3  - B_2 D - A B_1 - \Omega^2 /c^2 ) = 0.
\end{equation}
    The solution of Eq.~(\ref{tmS1}) is
    \begin{equation}    \label{sh2}
\Omega^2 = \frac{c^2 B}{2 r f - r^2 f'} \left( f'' r - 2r \frac{(f')^2}{f} + 3 f'
\right).
\end{equation}

    We can add linear functions satisfying Eqs.~(\ref{shx0}), (\ref{shx1}),
and (\ref{shx3}) to solutions~(\ref{shxRS}).
    Thus, the general solution of Eqs.~(\ref{shx0}), (\ref{shx1}), and (\ref{shx3})
for $\Omega^2 > 0$ is given by
    \begin{equation}    \label{OROTO1}
\left\{
\begin{array}{l}
\eta^{0} = \frac{ \sqrt{ \mathstrut 2r f'} }{f} \frac{ \omega }{ \Omega } \left(
C_1 \cos \Omega \tau - C_2 \sin \Omega \tau  \right) + C_{01} \tau  + C_{00}, \\[5pt]
\eta^r = C_1 \sin \Omega \tau + C_2 \cos \Omega \tau + C_{10}, \\[3pt]
\eta^\phi = \frac{ 2 }{r} \frac{ \omega }{ \Omega } \left(
C_1 \cos \Omega \tau - C_2 \sin \Omega \tau  \right) + C_{31} \tau + C_{30},
\end{array}
\right.
\end{equation}
    and for $\Omega^2 < 0$ we have
    \begin{equation}    \label{OROTOHip}
\left\{
\begin{array}{l}   \displaystyle
\eta^{0} = - \frac{ \sqrt{ \mathstrut 2r f'} }{f} \frac{ \omega }{ |\Omega| } \left(
C_1 \cosh |\Omega| \tau + C_2 \sinh |\Omega| \tau  \right) + C_{01} \tau  + C_{00}, \\[7pt]
  \displaystyle
\eta^r = C_1 \sinh |\Omega| \tau + C_2 \cosh |\Omega| \tau + C_{10}, \\[5pt]
  \displaystyle
\eta^\phi = - \frac{ 2 }{r} \frac{ \omega }{ |\Omega| } \left(
C_1 \cosh |\Omega| \tau + C_2 \sinh |\Omega| \tau  \right) + C_{31} \tau + C_{30},
\end{array}
\right.
\end{equation}
    where
    \begin{equation}    \label{OROTO1d}
C_{31} = - \frac{ B_1 C_{01} + c B_3 C_{10} }{ B_2 } ,
\end{equation}
and $ C_{1} $, $ C_{2} $, $ C_{00} $, $ C_{01} $, $ C_{10} $, and $ C_{30} $
are arbitrary constants.

    If we rename $C_1 \to C_1 r $, $C_2 \to C_2 r $, and $ C_{10} \to C_{10} r $,
then, in the case of the Schwarzschild metric, formulas~(\ref{OR2gen})
for $ \eta^\phi $ are reproduced at the first order in $r_g/r$.

    The general solution of Eqs.~(\ref{shx0}), (\ref{shx1}), and (\ref{shx3})
in the case $\Omega = 0$ can be found in terms of power functions:
    \begin{equation}    \label{OROTOom0}
\left\{
\begin{array}{l}  \displaystyle
\eta^{0} = - \frac{\sqrt{\mathstrut 2 r f'} }{3 f } \omega C_1 \tau^3 -
\sqrt{ \frac{r f'}{2}} \frac{ \omega}{f} C_2 \tau^2 + C_{01} \tau + C_{00}, \\[7pt]
\displaystyle
\eta^r = C_1 \tau^2 + C_2 \tau + C_{10}, \\[5pt]
\displaystyle
\eta^\phi =  - \frac{2 \omega}{3r} C_1  \tau^3 - \frac{\omega}{r} C_2  \tau^2 + C_{31} \tau + C_{30},
\end{array}
\right.
\end{equation}
    where
    \begin{equation}    \label{OROTO1p}
C_{31} = - \frac{ B_1 C_{01} + c B_3 C_{10} + 2 C_1/c^2 }{ B_2 } ,
\end{equation}
and $ C_{1} $, $ C_{2} $, $ C_{00} $, $ C_{01} $, $ C_{10} $, and $ C_{30} $
are arbitrary constants.

    Comparison of the obtained general solutions~(\ref{OROTO1}), (\ref{OROTOHip}),
and (\ref{OROTOom0}) with Eq.~(\ref{OR2gen}) results in the following
conclusion: the character of the time dependence of the circular reference
trajectory deviation is the same for the nonrelativistic and relativistic cases.
    The radial deviations from the circular reference trajectory take the form
of harmonic oscillations with the frequency~$\Omega $ in the case $\Omega^2 >0$.
    Such deviations grow quadratically with time for the limiting case $\Omega =0$
and exponentially for $\Omega^2 <0$ (for unstable circular orbits).
    The deviations in the angle~$\phi$ take the form of harmonic oscillations
at the frequency $\Omega $ with an additional linear growth in time in
the case$\Omega^2 >0$.
    Such deviations grow proportionally to the cube of time for the limiting
case $\Omega =0$ and exponentially for $\Omega^2 <0$
(for unstable circular orbits).

\vspace{3mm}
{\centering \section{\normalsize Shirokov effect and shift of orbit
pericenter \label{Sec4}}}

\hspace{\parindent}
    The difference between the frequencies $\Omega$ and $\omega$
leads to the shift of the pericenter of the perturbed orbit.
    The distance~$r$ to the center of gravity of the perturbed orbit is
determined by the value $\eta^1 $ and oscillates with the frequency~$\Omega$.
    During the time interval $ T= 2\pi/\omega$ of the body's rotation in
the unperturbed orbit, the shift of the pericenter is given by
    \begin{equation}    \label{SmPr}
\Delta \phi = 2\pi - T \Omega = 2 \pi \left(
1 - \frac{\Omega}{\omega} \right).
\end{equation}
     For instance, let us calculate the pericenter shift for an orbit close
to a circular one in rotation around a body of mass $M$ in the presence of
the cosmological constant~$\Lambda$.
    The corresponding interval represents the Kottler solution~\cite{Kottler18}:
    \begin{equation}    \label{mKot}
d s^2 = f(r)\, c^2 d t^2 - \frac{d r^2}{f(r)} - r^2 \left( d \theta^2 +
\sin^2 \theta \, d \phi^2 \right),
\end{equation}
    where
    \begin{equation}    \label{mKot2}
f(r) = 1- \frac{r_g}{r} - \frac{\Lambda r^2}{3},
\end{equation}
    and $r_g = 2 GM /c^2$.

    In this case, frequencies~(\ref{sh1}) and (\ref{sh2}) are given by
    \begin{equation}    \label{sh1K}
\omega^2 = \frac{c^2}{3 r^2} \cdot \frac{ 3r_g -2 \Lambda r^3}{ 2r - 3 r_g} ,
\end{equation}
    \begin{equation}    \label{sh1K1}
\Omega^2 = \frac{r_g c^2}{2 r^3} \cdot \frac{ 1- \frac{3 r_g}{r}-
\frac{8 \Lambda r^3}{3r_g} \left( 1 - \frac{15 r_g}{8 r} \right) }{
1 - \frac{3 r_g}{2 r}} .
\end{equation}

    The shift of the orbit pericenter for metric~(\ref{mKot}) during one
rotation in the first-order approximation in $r_g/r$ and $\Lambda r^3/r_g$ equals
    \begin{equation}    \label{sh1K2}
\Delta \phi = \pi \left( \frac{3 r_g}{r} + \frac{2 \Lambda r^3}{r_g}
\right) .
\end{equation}
    For $\Lambda =0 $ this formula reproduces the result for the perihelion
shift when the orbit eccentricity tends to zero
(see.~(13) in~\cite{Einstein} or (101,7) in~\cite{LL_II}).

    The difference between the frequencies $\Omega$ and $\omega$
of oscillations in mutually perpendicular directions for the motion of
a test body in a circular orbit in the Schwarzschild metric was calculated
by Shirokov~\cite{Shirokov73} using the geodesic deviation equation.
    This phenomenon is called the Shirokov effect~\cite{Nduka77},
\cite{Vladimirov81}, \cite{Ivanitskaya}, \cite{Vladimirov}.
    For the gravitational field of a point mass in the nonrelativistic case,
the frequencies $\Omega$ and $\omega$ were shown to be equal to each other.

    If the value of the orbit pericenter shift in a given gravitational field
is known, then passing to the limit of a circular trajectory gives the value of
the difference between the frequencies $\Omega$ and $\omega$.
    Therefore, this quantity for the Schwarzschild metric can also be obtained
as a consequence of the formula for the Mercury perihelion shift found by
Einstein~\cite{Einstein}.
    However, calculating the motion trajectory and the orbit pericenter
shift for other gravitational fields can be a complicated technical problem.
    Thus, the solution to the problem of the orbit pericenter shift for
the Kottler metric was obtained earlier in~\cite{HackmannLammerzahl08}.
    However, applying the general formulas from~\cite{HackmannLammerzahl08}
requires solving the complicated problem of the inversion of
ultraelliptic integrals (see the discussion in~\cite{Pavlov2025TMF}).

    Using the equations for geodesic deviation allows one to obtain
the values~(\ref{sh1}), (\ref{sh2}) of the oscillation frequencies~$\omega$
and $\Omega$ for perturbations of circular orbits directly from the metric,
providing an effective method even if the exact solution is known.

    As follows from~(\ref{sh1K2}), the influence of the cosmological constant
will be revealed in observations if the appropriate contribution to
the pericenter shift by the order of magnitude is no less than
the classical effect of general relativity, i.e.,
    \begin{equation}    \label{sh1K3}
\Lambda \ge \frac{r_g^2}{r^4} .
\end{equation}
    Thus, the cosmological constant for satellite orbits near the Earth
could be revealed in the effects of the orbit perigee shift for
values $\Lambda \ge 10^{-32}$\,m$^{-2}$.
    For the orbits of objects around the Sun, the effect of the cosmological
constant could be registered for values $\Lambda \sim 10^{-37}$\,m$^{-2}$
(for $r \sim 10^{11}$\,m).
    The observed value of the cosmological constant is much less than
those values, $\Lambda \sim 10^{-52}$\,m$^{-2}$.
    Its effects on the pericenter shift of the motion of Earth and Sun
satellites are far beyond the limits of experimental detection.
    However, if the accelerated expansion is driven by some scalar field,
namely quintessence~\cite{GorbunovRubakov}, then it is conceivable to
assume different values of quintessence in different regions.
    Then the data on the orbit pericenter shifts could serve as an indicator
of the value of this quintessence in the region near the Earth or
in the Solar System.

\vspace{3mm}
{\centering \section{\normalsize Metrics with frequencies
$\Omega = \gamma \omega$ \label{Sec5}}}

\hspace{\parindent}
    For $\gamma = m/n $ with natural $m$ and $n$, the oscillation frequencies
in mutually perpendicular directions are comparable and, consequently,
the trajectories are closed curves.
    We consider the case $B=f$.
    Then the requirement $\Omega = \gamma \omega$ reduces to the equation
    \begin{equation}    \label{OTOO}
3 f' f + f'' f r - 2r (f')^2  - \gamma^2 f' = 0 .
\end{equation}
    This equation has a solution given by
    \begin{equation}    \label{Van1}
f = \gamma^2 \cdot \frac{\frac{r^q}{2- q} + \frac{P}{ 2+q }\left( 4
- \gamma^2 \right)}{r^q + P \left( 4- \gamma^2 \right) } ,
\end{equation}
    where $P$ and $q$ are arbitrary constants.
    Next, we choose $q=2- \gamma^2 $ and assume that $\gamma^2 \ne 2$.
    In the case $ \gamma^2 < 2 $, such a metric is asymptotically flat
as $r \to \infty$.
    For the specified choice of~$q$, the $g_{00}$ component of
the metric equals
    \begin{equation}    \label{Van2}
f =  \frac{r^{2- \gamma^2} + P \gamma^2 }{r^{2- \gamma^2} +
P \left( 4-\gamma^2 \right) } =
1 - \frac{ 2 P \left( 2-\gamma^2 \right) }{ r^{2- \gamma^2} +
P \left( 4-\gamma^2 \right)  }.
\end{equation}

    In the nonrelativistic approximation (see~\cite{LL_II})
the following potential~$ \varphi $ corresponds to the metric with
such a $00$-component:
    \begin{equation}    \label{Van3LL2}
f = 1 + \frac{2 \varphi}{c^2} ,
\end{equation}
    \begin{equation}    \label{Van3}
\varphi = \frac{ - P c^2 \left( 2-\gamma^2 \right) }{ r^{2- \gamma^2} +
P \left( 4-\gamma^2 \right) }.
\end{equation}
    If $ \gamma^2 \ne 2 $, then attraction occurs for $P>0$.

    The trajectories close to circular orbits in central fields in general
relativity are closed if the metric interval is of form~(\ref{Van2}) with
    \begin{equation}    \label{ZamOto}
\gamma = \frac{m}{n} , \ \ \ m,n \in \mathbf{N} .
\end{equation}

    If $ r^{2- \gamma^2} \gg P ( 4-\gamma^2 ) $, then formula~(\ref{Van3})
reproduces the expression for the potentials with closed deviations of
circular orbits in the nonrelativistic case~(\ref{sch4}).
    If $\gamma = 2 $, then we obtain the metric of the anti-de Sitter space
with the cosmological constant $\Lambda = -12 P $:
    \begin{equation}    \label{Van4}
f =  1 + 4 P r^2 = 1- \frac{\Lambda r^2}{3},
\end{equation}

    Solutions~(\ref{Van2}) are regular as $r \to 0$:
    \begin{equation}    \label{Van5}
r \to 0 \ , \ \gamma^2<2 \ \ \Rightarrow \ \ f \to  \frac{\gamma^2}{4- \gamma^2},
\end{equation}
    \begin{equation}    \label{Van6}
r \to 0 \ , \ \gamma^2 \ge 2 \ \ \Rightarrow \ \ f \to  1.
\end{equation}

    If $\gamma >2$, then, for the metric with components $g_{00} = - 1/g_{rr}=f$,
the singularity at $ r = (P(\gamma^2 - 4) )^{-1/(\gamma^2 -2)}$
is a coordinate one rather than a spacetime singularity.
    The determinant of the metric components is as follows:
${\rm det} \{ g_{ik} \} = - r^4 \sin^2 \theta \ne 0$ for $\theta \ne 0, \pi$
and for this value of~$r$.

\vspace{3mm}
{\centering \section{\normalsize Metric with $\Omega = 0$
\label{Sec6}}}

\hspace{\parindent}
    The zero value of $\Omega$ corresponds to the limiting case of stable
circular orbits.
    In classical mechanics, the potential $ \sim 1/r^2$ corresponds to this case.
    For $\Omega=0$, within general relativity, we derive the following equation
for the metric components from Eq.~(\ref{sh2}):
    \begin{equation}    \label{OTOOm0}
3 f' f + f'' f r - 2r (f')^2  = 0 .
\end{equation}
    To find the general solution of this equation, we introduce
the substitution $z = f'/f$.
    Then equation~(\ref{OTOOm0}) reduces to
    \begin{equation}    \label{Va1}
3 z f^2 + r (z'+z^2) f^2 - 2r z^2 f^2  = 0 .
\end{equation}
    For $f\ne 0$, we obtain the Riccati equation
    \begin{equation}    \label{Va2}
z' = -\frac{3}{r} z + z^2 .
\end{equation}
    The substitution $u=1/z$ leads to the linear equation
    \begin{equation}    \label{Va3}
u' -\frac{3}{r} u = - 1 .
\end{equation}
    The solution of this linear equation is given by
    \begin{equation}    \label{Va4}
u = \frac{1}{2} r + C_1 r^3 ,
\end{equation}
    where $C_1$ is an arbitrary constant.
    Returning to the function~$f$, we find that the general solution of
Eq.~(\ref{OTOOm0}) can be written as follows:
    \begin{equation}    \label{OTOTR}
f = \frac{\gamma r^2}{r^2 + a^2},
\end{equation}
    where $\gamma $ and $a$ are arbitrary constants.
    If we require the spacetime to be asymptotically flat at infinity
($f \to 1$ as $r \to \infty$), we must set $\gamma = 1 $:
    \begin{equation}    \label{Va5}
f =  1 - \frac{a^2}{r^2 + a^2} ,
\end{equation}
    where $a^2>0$ in the case of attraction, when circular orbits are possible.

    In the nonrelativistic approximation, the potential
$ \varphi \sim - c^2 a^2 /(2 r^2) $ corresponds to such a solution
as $r/a \to \infty $.

\vspace{3mm}
{\centering \section{\normalsize Metric without Shirokov effect
\label{Sec7}}}

\hspace{\parindent}
    In the case of Eq.~(\ref{OTOO}) with $ \gamma = 1 $, solution~(\ref{Van2})
is given by
    \begin{equation}    \label{Van7}
f = 1 - \frac{ 2 P }{ r + 3 P } ,
\end{equation}
    where the constant $P>0$.
    For such a static spherically symmetric spacetime,
$ \Omega = \omega $ and there is no Shirokov effect within general relativity.
    We show that the matter generating such a gravitational field does
not violate the energy conditions~\cite{HawkingEllis}.

    The components of the energy-momentum tensor of the matter producing
the gravitational field can be found from the Einstein equations:
    \begin{equation}
R_{ik} - \frac{1}{2} g_{ik} R = \frac{8 \pi G}{c^4}\, T_{ik} .
\label{GR70Ein}
\end{equation}
    For the centrally symmetric static metric~(\ref{msf}),
according to~\cite{LL_II} (see~\S~100 in that book), we obtain the
following nonzero components of the energy-momentum tensor:
    \begin{equation}
\frac{8 \pi G}{c^4}\, T_{0}^0 = \frac{1}{r} \left( \frac{1 - B}{r} - B'\right),
\label{T00TS}
\end{equation}
    \begin{equation}
\frac{8 \pi G}{c^4}\, T_{1}^1 = \frac{1}{r} \left( \frac{1 - B}{r} -
\frac{B f'}{f}\right),
\label{T11TS}
\end{equation}
    \begin{equation}
\frac{8 \pi G}{c^4}\, T_{2}^2 = \frac{8 \pi G}{c^4}\, T_{3}^3 =
- \frac{B}{2} \left( \frac{f''}{f} - \frac{1}{2} \left( \frac{f'}{f} \right)^2
+ \frac{1}{r} \left( \frac{f'}{f} + \frac{B'}{B} \right) +
\frac{f'B'}{2fB}\right) .
\label{TaaTS}
\end{equation}
    These nonzero components correspond to the energy density and pressure:
    \begin{equation}
T^i_k = {\rm diag}\, ( \varepsilon, -p^r, -p^\theta , -p^\phi\,).
\label{TilEP}
\end{equation}
    As we see from formulas~(\ref{T00TS}) and (\ref{T11TS}), if $f(r) = B(r)$,
then $ \varepsilon = -p^r $.
    The converse is also true: the requirement $ \varepsilon = -p^r $
leads to the equation $f(r) = {\rm const}\cdot B(r)$, which is equivalent to
the condition $f(r) = B(r)$ after redefining the time variable
$t' = t / \sqrt{ {\rm const} }$ in interval~(\ref{msf}).

    We have for metric~(\ref{Van7}):
    \begin{equation}
T_{0}^0 =  T_{1}^1 = \frac{c^4}{8 \pi G}\, \frac{ 6 P^2 }{r^2 (r+3 P)^2 },
\label{T00TSes}
\end{equation}
    \begin{equation}
T_{2}^2 = T_{3}^3 = - \frac{c^4}{8 \pi G}\, \frac{ 6 P^2 }{r (r+3 P)^3 }.
\label{T00TSes2}
\end{equation}
    According to~\cite{HawkingEllis} (\$~4.3), the energy conditions
(the weak energy condition and the dominant energy condition) hold for this
energy-momentum tensor: $\varepsilon \ge 0$ and $|p^i| \le \varepsilon $.

\vspace{4mm}
{\centering \section{\normalsize Conclusions
\label{secZakl}}}

\hspace{\parindent}
    In this paper, we have considered the deviation of circular trajectories
in the nonrelativistic case and in general relativity.
    For comparison, the general solutions~(\ref{ndev3d}), (\ref{OR1}),
and~(\ref{OR2gen}) of the deviation equations for central potentials are given
for the cases of stable orbits, the limiting case of a stable orbit,
and unstable circular orbits.
    We have found all centrally symmetric potentials~(\ref{sch4}) for which
trajectories close to circular ones are closed curves.

     We have presented the exact solutions of the deviation equations for
circular geodesics in static spherically symmetric spaces~(\ref{msf})
in general relativity: for stable orbits~(\ref{OROTO1}),
unstable orbits~(\ref{OROTOHip}),
and limiting stable circular orbits~(\ref{OROTOom0}).
    Comparison of the corresponding formulas leads to the conclusion that the
character of the time dependence of circular-orbit deviations is the same in
both the nonrelativistic and relativistic cases.

    We have shown that the difference between the trajectory oscillation
frequencies in different directions (the Shirokov effect) leads to the shift
of the pericenter of the perturbed orbit.
    Explicit expressions for the pericenter shift are given in Eq.~(\ref{sh1K2})
for the case of the metric of a gravitational mass in the presence of
the cosmological constant (the Kottler metric~(\ref{mKot})),
and possible observational evidence of this effect has been analyzed.

    We find the metrics (\ref{Van2}) and (\ref{ZamOto}) of the static
spherically symmetric spaces~(\ref{msf}), where geodesics close
to circular ones remain closed curves.
    We find metric~(\ref{Va5}), which represents the limiting case for stable
circular orbits in general relativity in a static spherically symmetric
spacetime.

    We find an example of a static spherically symmetric metric~(\ref{Van7})
in which the oscillation frequencies of the deviation vector for
circular orbits are the same in all directions.
    We find the components of the energy-momentum tensor~(\ref{T00TSes})
and~(\ref{T00TSes2}) of the matter generating the corresponding gravitational
field and show that the standard energy conditions hold for such matter.
    Thus, the Shirokov effect (the occurrence of different oscillation
frequencies of the deviation vector in different directions) can be absent
within general relativity for a certain distribution of matter in space.

\vspace{4mm}
{\bf Funding.}\,
    The work of Yu. V. Pavlov was performed within the framework of the state
assignment of the Ministry of Science and Higher Education of
the Russian Federation for IPMash RAS (topic No. 124041500009-8).

\vspace{4mm}
{\bf Conflict of interest.}\,
The authors of this work declare that they have no conflicts of interest.

\vspace{2mm}

\end{document}